# THERMAL CONDUCTIVITY THROUGH THE NINETEENTH CENTURY

T. N. Narasimhan[1]
May 11, 2010

*As a material property and as a metaphor, thermal conductivity occupies an important position in physical, biological and geological sciences. Yet, its precise measurement is dependent on using electricity as a proxy because flowing heat cannot directly be measured.*

## Eighteenth Century Beginnings

Following Joseph Black's (1728-1799) introduction of the concepts of latent heat and specific heat in 1761, a major advancement in thermal science was the investigation of heat conduction in solids by Johann Heinrich Lambert (1728-1777), who considered a long metal rod heated at one end, with heat allowed to dissipate to atmosphere by radiation. A similar experiment had originally been used by Guillaume Amontons (1663-1705), who assumed that temperature varied linearly along the rod. Lambert revisited the experiment so as to correct Amonton's assumption of linear temperature profile. The experiment and its interpretation were posthumously published two years later in 1779[1]. Lambert exhibited remarkable insight in framing steady-state heat conduction in terms of energy balance thus,

> "*...the heat flows gradually to the more distant parts, but at the same time travels from each part to the air. So that when the fire has burnt and been maintained long enough at the same strength, every part of the bar finally acquires a definite degree of heat because it constantly acquires as much heat from parts of the rod nearer the fire as transmits to the more distant parts and the air.*" (Ref 2, page 163).

He showed that the temperature profile along the rod declined logarithmically (Figure 1). He was insightful in recognizing the importance of geometry of the rod in governing the temperature profile. He observed (Ref 1, p. 187),

---

[1] Department of Materials Science and Engineering, Department of Environmental Science Policy and Management, University of California, Berkeley, Ca 94720-1760   Tnnarasimhan@LBL.gov

> *"...there is an uncertainty because Amontons does not say what his rod looked like .."*.

Lambert's conceptual framework later played an important role in Fourier's formulation of the heat equation.

About this time, Benjamin Franklin (1706-1790) had conceived a scheme to evaluate relative heat-conducting abilities of different metals by coating them with wax, heating them and observing the distance to which wax would melt in different cases. During a visit to France in 1780, he gave Jean Ingen-Housz (1730-1799) , a Dutch biologist and chemist, his experimental concept and materials he had collected, and encouraged him to conduct the experiments at his leisure.

Ingen-Housz coated a number of wires with wax, each of different material but of same length and diameter, by dipping them in molten wax and letting it cool and solidify. The wires were tightened in parallel between blocks of wood. He then dipped one end of the wires in hot oil, all wires at the same time and over the same length. He observed that the wax coat melted along all wires, but the speed of the melt propagation varied directly with the speed with which heat ran through these metals (Ref. 3, p.387). In all, he conducted twelve experiments using 7 metals. Results from one of these experiments is shown in Figure 2, with each metal being identified by its chemical symbol. The horizontal marks represent the distances to which wax had melted along different metals. Although there were variations, he found the following order of decreasing conductivity: silver, copper, gold, tin, iron, steel and lead.

The next major contribution was that of Count Rumford (1753-1814), who was motivated by a desire to evaluate the insulating properties of natural and artificial clothing. His basic apparatus (Figure 3) consisted of a thermometer enclosed and hermetically sealed within a carefully blown glass bulb. The bulb was filled with various materials of interest: torricellian vacuum, air, water, mercury, silk, wool, fur, and so on. The procedure was to plunge the apparatus first into a hot bath, then into freezing water and observing the gradual decline in temperature as a function of time, or reversing the process by first plunging the apparatus into freezing water and then into boiling water. Regardless, the time-rate of loss or gain of heat as revealed by the temperature-time relation was considered a measure of thermal conductibility. Based on his experiments, Rumford (Ref. 4, p. 83) concluded that the relative conducting powers of materials were as follows: mercury (1000), moist air (330), water (313), common air (80.41), rarefied air (80.23), and torricellian vacuum (55).

By the close of eighteenth century, thermal conductivity had been understood in an intuitive sense, but imprecisely. Both Ingen-Housz and Rumford were unable to recognize that the time-rate of change of temperature depends on a combination of thermal conductivity and specific heat. Lambert astutely focused on the steady-state. In his experiment, thermal conductivity was masked by effects of heat loss by radiation. Regardless, the experimental methods of Lambert and Ingen-Housz would have significant influence on heat conduction investigationsduring the nineteenth century.

**Fourier's Theory of Heat Conduction**

In 1802, upon return to France from Napoelon's Egyptian campaign, Jean Baptiste Joseph Fourier (1768-1830) was appointed prefect of the department of Isère. Despite heavy administrative responsibilities, Fourier chose to devote personal time to study heat diffusion[5]. In this, he was inspired by deep curiosity about the earth: attenuation of seasonal temperature variations in the earth's subsurface, oceanic and atmospheric circulation driven by solar heat, and background temperature in deep space[6]. Fourier started with a paper by Jean Baptiste Biot (1774-1862)[7]. Without citing Lambert, Biot attempted to formulate a differential equation for heat conduction in a rod heated at one end with heat allowed to dissipate to atmosphere[2,5]. His approach, based on action-at-distance and Newton's Law of cooling was unsuccessful because Newton's Law, appropriate for radiative heat loss, was inadequate for conductive transfer.

Starting with Biot's approach, Fourier obtained some mathematical results that were incorrect and unsatisfactory. He then abandoned action-at-distance, and devoting attention to the physical phenomenon he recognized that temperature varied continuously along the length of the rod (Figure 4). Also, conservation of heat, as conceived by Lambert, required that conductive heat flow along the rod be balanced by transverse radiative heat-loss to atmosphere. Accordingly, he made a distinction between external conduction, governed by Newton's Law, and internal conduction. For the latter, he defined thermal conductivity as the quantity of heat per unit time passing through a unit cross section of a prism per unit difference in temperature between two isothermal surfaces[8]. He treated linear temperature variation (Figure 4) implicit in this definition as being synonymous with a gradient:

$$-K[\frac{B-b}{A-a}] = -K\frac{\Delta y}{\Delta x} \qquad (1)$$

Nevertheless, this development presented in the 1807 unpublished memoire did not satisfy Fourier. It took him another three years to go beyond the discrete finite-difference description of flow between isothermal surfaces and express heat flux across an inifinitesimally thin surface segment in terms of temperature gradient. This enabled him to express the boundary condition at the radiative surface per unit area per unit time as, hT = -K∂T/∂**n**, where h is a heat transfer coefficient, T temperature, and **n** is unit outer normal. A detailed discussion of this transition can be found in Grattan-Guinness[5] and Herivel[2]. In Fourier's final formulation, the flow-process in the interior was made distinct from that at the surface.

**Fourier's Experiments**

At the turn of the nineteenth century, there was debate as to whether heat was a fluid (caloric) or a kind of motion. Nevertheless, Fourier[6] considered mathematical laws to which effects of heat were subject to be independent of all hypothesis about the nature of heat. It was known that the quantity of heat contained in a body at a given temperature could be quantified using an ice calorimeter invented by Antoine Lavoisier (1743-1794) and Pierre Simon Laplace (1749-1827)[9] in terms of quantity of heat required to change the state of a certain mass of ice to liquid. But, no method was available to measure flowing heat. Consequently, Fourier had to devise suitable experiments and methods to measure thermal conductivity so that his mathematical theory could be credible.

It is not widely recognized that in his unpublished 1807 memoire and in the Prize Essay submitted in 1811[10], Fourier provided results from transient and steady-state experiments, and outlined methods to invert experimental data to estimate thermal conductivity. For some reason, he decided to restrict his 1822 masterpiece purely to mathematics, omitting experimental results.

In all, Fourier carried out four sets of experiments[5]: 1) distribution of temperature in a heated annulus, 2) rate of cooling of an annulus, 3) rate of cooling of a sphere, and 4) comparison of rates of cooling of sphere and cube. Of these, (1) and (3) are pertinent to thermal-conductivity determination.

The first experiment consisted of a polished iron annulus, about 0.35 meters in diameter , with a square cross section, heated at one point with an Argand lamp. Six holes had been half drilled through the annulus, in regular spacing on opposite quadrants to receive thermometers.  Heat was allowed radiate to atmosphere across the polished surface.  Mathematically, steady-state heat flow in the annulus was governed by the ordinary equation describing Lambert's experiment, namely,

$$\frac{d^2T}{dx^2} = \frac{2h}{KL}T, \tag{2}$$

where T is temperature, x distance along abcissa, h external conducting coefficient, K thermal conductivity, and 2L the side of the square cross section of the annulus.  The solution to this equation is given by[7],

$$T = T_0 e^{-x\sqrt{\frac{2h}{KL}}}, \tag{3}$$

where $T_0$ is T(x = 0).  This result effectively confirms Lambert's empirical suggestion (Figure 1) that temperature profile would be logarithmic.  Based on this result, Fourier found that if three thermometers were spaced equidistant, then, the quotient $(T_1 + T_{3)//}T_2$  should be a constant, which was confirmed by his experimental results.

The other experiment involved a small sphere of radius X with constant temperature $T_0$ in the interior at t = 0, and  allowed to cool in an atmosphere maintained at constant temperature, with h << K.  For this case, Fourier found that temperature inside the sphere would vary very little around an average value

$$T = T_0 e^{-\frac{3ht}{CDX}}, \tag{4}$$

where C is specific heat and D is solid density.

In addition, one other result needs mention.  In Part II of the Prize essay[10] , Fourier addressed a terrestrial heat problem that had inspired him, namely, propagation of diurnal and seasonal temperature changes within the earth's crust. Considering that the earth's diameter is very large, he idealized an infinite one-dimensional vertical column extending down from the land surface (x = 0), with temperature everywhere being zero at the initial time, subjected to a periodic variation of temperature at x = 0 for t > 0. The periodic function, in general, is a sum of many sine and cosine waves. For a simple periodic temperature signal at the land surface with mean $T_0$, amplitude A, period $\theta$, and phase $\varepsilon$, he showed that the heat equation is satisfied by,

$$T(x,t) = T_0 + A\, e^{-x\sqrt{\frac{\pi CD}{K\theta}}} \sin\left(\frac{2\pi t}{\theta} + \varepsilon - x\sqrt{\frac{\pi CD}{K\theta}}\right) \;. \tag{5}$$

Using this result, he deduced that with increasing depth, temperature amplitude will attenuate and phase will shift with depth x, both governed by $\sqrt{\frac{\pi CD}{K\theta}}$ .

**First Determination of Thermal Conductivity**

The first experimental determination of thermal conductivity of a solid material was made by Fourier (Ref 10, p. 14-15) for steel. In the absence of any method for measuring flowing heat, he had to rely on known magnitude of external conductivity h and change in heat content of the material to estimate flowing heat. Accordingly, using (3) which represents T(x) for the annulus problem, he estimated h/K from the relation, $\dfrac{\ln\left(\dfrac{T_1}{T_2}\right)}{x_2 - x_1} = \sqrt{\dfrac{2h}{KL}}$ .

Next, by applying (4) to data from the sphere experiment, he estimated h/(CD), by the relation,

$$\frac{\ln\dfrac{T_1}{T_2}}{t_2 - t_1} = \frac{3h}{CDX} \;.$$

Assuming that volumetric specific heat CD is independently known, he combined the known estimates of h/K and h/CD to arrive at a value of 3/2 for thermal conductivity in units of meter, minutes and kilogram, with units for the quantity of heat being the quantity required to convert one kilogram of ice at 0 degrees to an equal mass of water at the same temperature.

Then, drawing upon (5), the solution to the periodic boundary condition problem, Fourier substituted the known value of CD/K into $\sqrt{\dfrac{CD\pi}{K\theta}}$ ,

and using θ = 1,440 minutes, showed that the amplitude of the boundary signal would be negligible at depth of about 2.3 meters. This confirmed the prevalent knowledge that diurnal temperature variations die out within a few meters of land surface.

**Rest of the nineteenth century**

With solutions (3), (4), and (5), and by providing the first estimate for thermal conductivity of steel, Fourier had set the conceptual-mathematical stage for investigations for the rest of the nineteenth century. In the absence of a method for directly measuring flowing heat, research on experimental determination of thermal conductivity was pursued along two broad lines. The first was to quantitatively establish the relative conductivities of different materials, and the second was to estimate absolute conductivity of specific materials by inverting experimental data.

*Relative Conductivity*

Recall that Ingen-housz used the rapidity with which melting progressed in different materials as a measure of relative conductivity. This reasoning is not entirely correct because, as shown by Fourier in (4), time-rate of diffusion depends on the quotient K/CD, which is now known as thermal diffusivity. Accordingly, César-Mansuète Despretz (1798-1863) ventured to determine relative thermal conductivities more precisely using Fourier's solution (3) to Lambert's problem. For three thermometers spaced uniformly along a rod, Fourier had shown that $(T_1 + T_3)/T_2$ is a constant. Let 2C be such a constant, and consider two materials A and B having identical dimensions and exterior conductance h. It can be shown[11] that their relative conducting ability is given by,

$$\sqrt{\dfrac{K_A}{K_B}} = \dfrac{\ln(C_B + \sqrt{C_B^2 - 1})}{\ln(C_A + \sqrt{C_A^2 - 1})} \ . \tag{6}$$

Using this approach, Despretz found that the conductivity of copper is larger than that of iron by a ratio of 12 to 5. His experiments indicated the following order, from the most conducting to the least conducting material: copper, iron, zinc, tin, lead, marble, porcelain and brick.

*Absolute conductivity*

The most direct method of determining absolute thermal conductivity of a material would be to take a slab of material with parallel walls, with one wall being maintained at boiling point of water by continuous supply of steam, and the other maintained at zero degrees by continuous melting of ice. Then, the flux of heat between the walls could be estimated from the quantity of steam condensed or the quantity of ice melted. First introduced by Jean Claude Peclet (1793-1857)[12] this approach and its variants were attempted by others, with limited success.

This left two approaches, both suggested by Fourier. The first was to combine steady-state data from a rod heated at one end (Lambert's experiment), with transient cooling of a segment of the same of the material in the atmosphere. The second approach was to impose a periodic temperature variation at the boundary and recording the attenuation of amplitude or lagging of phase with distance from the boundary.

James David Forbes (1809-1868)[13] determined the absolute conductivity of a 8-foot long bar of wrought iron with a 1.25-inch square cross section using Lambert's experimental design, supplemented by a second experiment in which a segment of the same bar, provided with a thermometer, was allowed to cool in air under the same conditions as the other experiment. Note that in determining the thermal conductivity of steel in 1811, Fourier[10] had used a similar two-step procedure, the first step providing the ratio h/K and the second providing the ratio h/CD. Forbes departed from Fourier's approach of estimating ratios, and chose to use a graphical procedure. Accordingly, he plotted temperature T in excess of atmospheric temperature as a function of x from the steady-state experiment, and fitted a smooth curve to enable graphical estimation of T and dT/dx at any point. He also plotted the decline of T with time t from the second experiment and fitted a curve that enabled graphical estimation of dT/dt and the corresponding temperature.

Forbes recognized that the steady quantity of heat -K(dT/dx) crossing the bar at x will be exactly equal to the quantity of heat lost to the atmosphere from the bar beyond x. Thus, by estimating dT/dx from the first curve, and by estimating the quantity of heat lost to the atmosphere using the second curve, K could be estimated. Consider the vicinity Δx on the bar with temperature T. Corresponding to this T, let dT/dt be the rate of decline as read from the experimentally determined curve. Then, the time-rate at which heat is lost to atmosphere is given by (CDAΔx)dT/dt, where A is the area of cross section. Thus, if the bar beyond x is divided into I segments, the total heat-loss ΔH to atmosphere is,

$$\Delta H = \sum_i^I CDA\Delta x \left(\frac{dT}{dt}\right)_i . \qquad (7)$$

Then, K = (ΔH)/ (dT/dx). Forbes' tabulated interpretation showed that thermal conductivity for wrought iron decrease from 0.107 at 100° C to 0.0082 at 200° C, a decline of about 23%. Forbes' graphical procedure established for the first time that absolute conductivity of wrought iron depends on temperature.

Forbes (Ref. 13, p. 136) states that as far as he knew no other method had succeeded in measuring absolute conductivity. Clearly, he was unaware of Fourier's Prize Essay[10].

William Thomson (Lord Kelvin ,1824-1907) used the solution to the heat equation under a periodic boundary condition to decipher thermal conductivity of the earth from underground temperature measurements made over a period of 13 years by Forbes at Edinburgh. Kelvin too was likely unaware of Fourier's Prize Essay[10] in which the required solution (5) had been presented. Instead, he relied on an independent derivation showing that the attenuation of the logarithm of amplitude (Δ ln A) or phase-lag (ΔE) between two planes $x_1$ and $x_2$ ($x_2 > x_1$) is,

$$\frac{\Delta \ln A}{x_2 - x_1} = \frac{\Delta E}{x_2 - x_1} = \sqrt{\frac{\pi CD}{\theta K}} . \qquad (8)$$

By analyzing long-term temperature observations at different depths, Kelvin estimated thermal conductivity of trap rock, sand, and sandstone.

**Directional Dependence in Crystals**

Henri Hureau de Sénarmont (1808-1862), a mineralogist, conducted experiments on thin,

oriented plates of various crystals (and bodies subjected to pressure in one direction) with a small hole drilled in the center, coating them with wax, and continuously applying heat through a heated wire or a tube inserted into the hole. He observed the shape of the isothermal envelope radiating out from the point as evidenced by the melting wax. He found that in crystals of cubic system or amorphous materials, the envelope was circular. But, in other systems the envelope had an elliptic or oval shape, depending on orientation. For planes parallel to the c axis in quartz, for example, the major axis of the ellipse was 1.31 times that of the minor axis[15]. Sénarmont reported his results, and qualitatively established that thermal conductivity, just as refractive index, is directionally dependent, with principal directions coinciding with crystallographic axes.

George Gabriel Stokes (1819-1903) provided a mathematical theory to explain Sénarmont's observations[16]. Treating the crystal as a continuum with different conductivities along three perpendicular directions, he derived expressions to calculate heat flux across an area-segment whose orientation relative to crystallographic axes is known, along with the temperature gradient normal to the surface. The matrix of thermal conductivity components resulting from his derivation satisfied certain invariance criteria, and hence constituted a tensor. This directional dependence later came to be known as anisotropy.

**Epilogue**

An important contribution to 19$^{th}$ century physics was the experimental discovery by James Prescott Joule (1818-1889) of the law that the quantity of heat generated per unit time during propagation of voltaic electricity in a conductor is proportional to the product of the resistance of the conductor and the square of the current. For the first time, this Law provided a way of estimating the quantity of electrically generated flowing heat. Nevertheless, till the very end of 19$^{th}$ century, precise measurement of thermal conductivity was impeded by having to indirectly estimating flowing heat from surface emissivity or specific heat. Among the earliest to introduce electrical heaters as controlled sources for heat in thermal conductivity measurements was Charles Lees[17].

In retrospect, one is impressed by Lambert's grasp of the significance of Amonton's experiment, and expanding on it based on energy conservation. Fourier's powerful theory had to have credibility when no known method existed for measuring flowing heat. Here, external conductivity provided a way of quantifying flowing heat as Fourier demonstrated in experimentally determining thermal conductivity for the first time in his prize essay[10]. For the same reason, Lambert's experimental design was also latent in the investigations of Despretz and Forbes.

Fourier's Analytic Theory of Heat, published in 1822, is considered by many to be among the most important contributions of modern science. For some reason, Fourier did not include in this masterpiece his experimental verification of theory, the two-step procedure to experimentally determine thermal conductivity, and the solution to the heat equation subject to a periodic boundary condition. These were presented in his unpublished memoir of 1807, and the prize essay of 1811. Although the latter would be published subsequently in Mémoires de l'Academie Royale des Sciences in 1826, it escaped the attention of distinguished scientists such as Forbes and Kelvin. Thus, in retrospect we find that Fourier had in fact set the stage for the methods of measuring thermal conductivity throughout the nineteenth century.

Thermal conductivity plays a very important role in the history of physics. Fourier struggled for several years with giving it a satisfactory conceptual-mathematical description, as he was boldly breaking away from action-at-distance and planetary mechanics to set in place the continuum paradigm. Thermal conductivity, as defined by Fourier not only was a physical property of fundamental importance, but also became a metaphor that inspired defining other physical properties of importance such as electrical resistivity, coefficient of molecular diffusion, and flow of fluids in resistive media. Intrinsic to Onsager's reciprocal relations, these coefficients are widely used in physical, biological and geological sciences. Yet, it is remarkable that flowing heat continues to elude our ability to measure directly.

**Acknowledgments**

I would like to thank Karsten Pruess and Nic Spycher for help with German and French translation. Work was partly supported by Committee on Research, University of California, Berkeley.

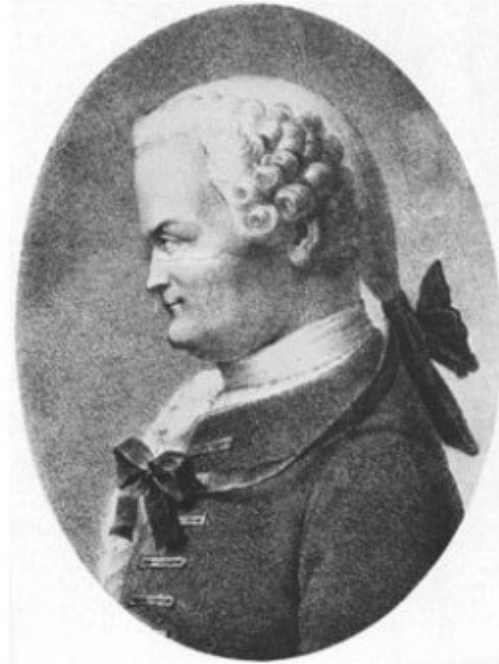

Figure 1: Johann Heinrich Lambert was a versatile scientist and philosopher. He formally introduced the heat-conservation framework for analyzing heat conduction in a rod heated end and allowed to radiatively dissipate heat to atmosphere. He showed that temperature y in his experimental data satisfied an exponential relation, $y = 97 + 1065\, e^{-\frac{x}{116.3}}$ , where x is distance from the heated end (Table on Left). In the Table, column 1 shows distance, columns 2 and three show temperatures, respectively, in Fahrenheit and Newton scales. These temperatures were based on various substances (column 4) that were observed to melt at the corresponding distances. From top to bottom, these materials are, white-hot iron, thin glass, lead, tin, lead-tin alloy, boiling water, wax, melting tallow, and butter.

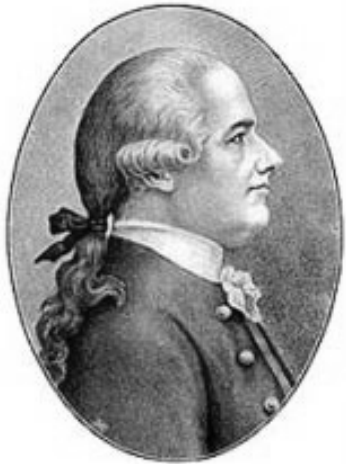 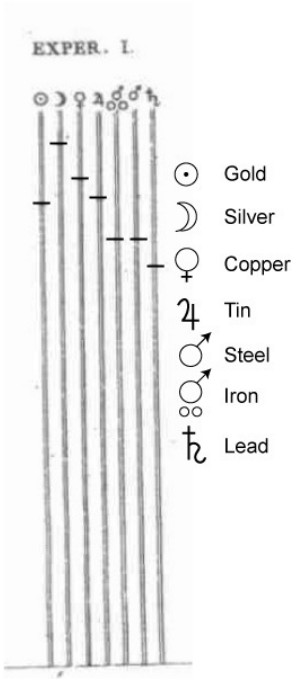 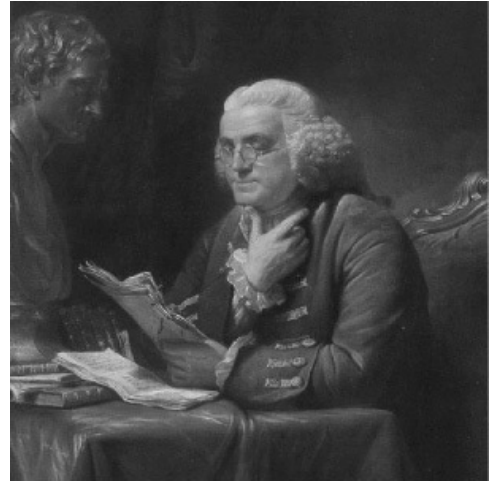

Figure 2: In a trans-Atlantic collaboration, Jean Ingen-Housz (Left) evaluated relative conductivities of different metals based on Benjamin Franklin's (Right) experimental design and materials. In gracious acknowledgment, Ingen-Housz stated, "...If these results bring light to the question we are trying to resolve, it is to this great man that physics will be indebted, not to me, who only followed exactly the project of Mr. Franklin." (Ref. 3, p. 386). Six decades later, this method helped establish directional dependence of thermal conductivity in crystals

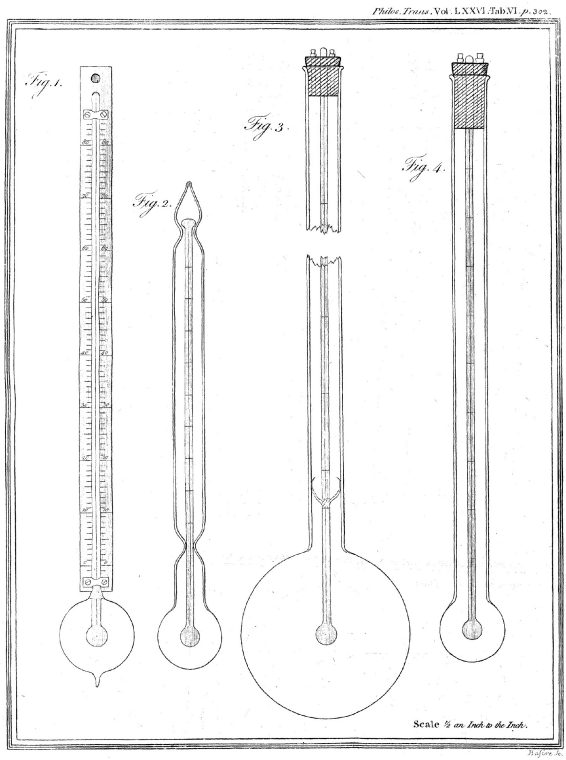 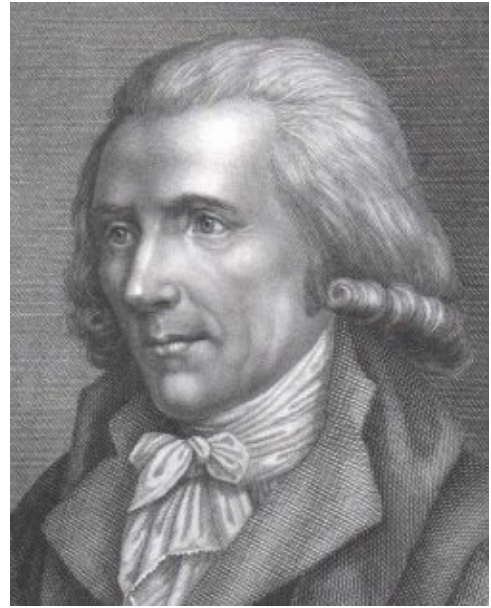

Figure 3: Count Rumford was a scientist, inventor and an entrepreneur. He conducted experiments on heat conduction "... with a view to the investigation of the causes of the warmth of natural and artificial clothing" (Ref 4). He enclosed thermometers in carefully blown glass bulbs filled with various materials, and used the time-rate of cooling as a measure of their "conducting power".

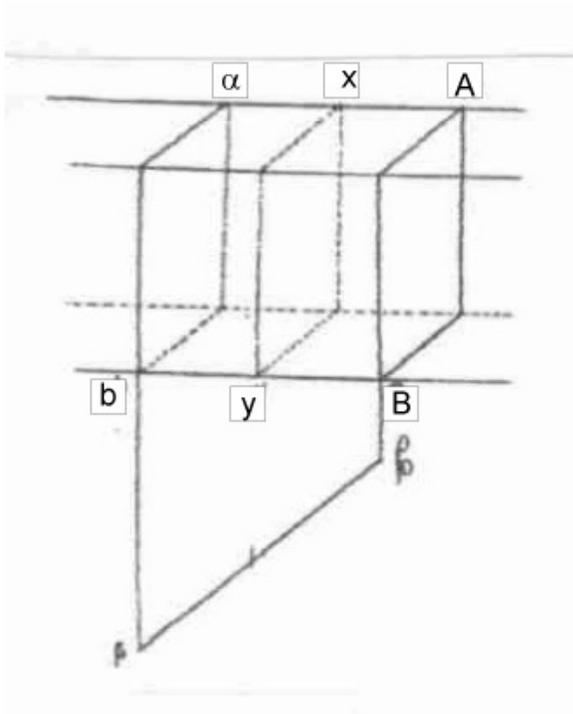 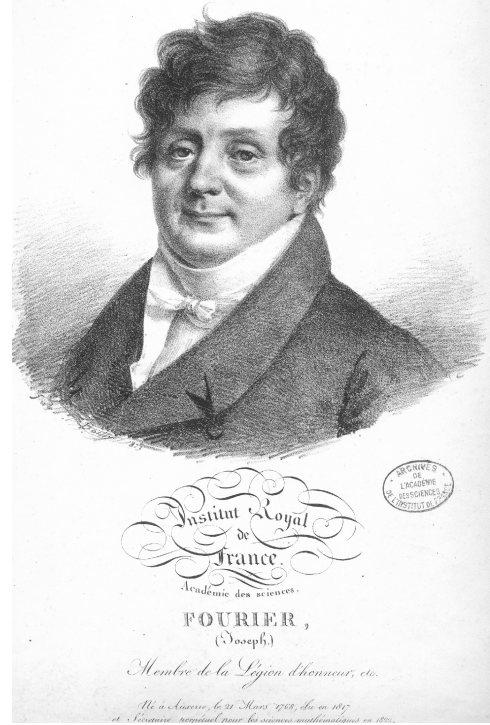

Figure 4: Fourier broke away from action-at-distance among disconnected bodies, the prevalent mind-set in Newtonian physics, and conceived of continuous, steady flow of heat in a prism of uniform cross section (Left). This step, which led to a physical-mathematical definition of thermal conductivity played a pivotal role in the development of his theory of heat.